\documentstyle[epsf,aaspp4,flushrt]{article}

\newcommand{\simgt}{\lower.5ex\hbox{$\; \buildrel > \over \sim \;$}}
\newcommand{\simlt}{\lower.5ex\hbox{$\; \buildrel < \over \sim \;$}}
\newcommand{\citet}[1] {\cite{#1}}
\newcommand{\citep}[1] {(\cite{#1})}
 \newcommand{\bm}[1]{\mbox{\boldmath$#1$}}
 \newcommand{\kaco}[1]{\left\langle{#1}\right\rangle}
 \newcommand{\skaco}[1]{\langle{#1}\rangle}
 
 \newcommand{\solam}{M_\odot}
\newcommand{\rvir}{r_{\rm vir}}
\newcommand{\rcor}{r_{\rm c}}
\newcommand{\baredth}{\;\overline{\raise1.0pt\hbox{$'$}\hskip-6pt
\partial}\;}
\newcommand{\edth}{\;\raise1.0pt\hbox{$'$}\hskip-6pt\partial\;}
\begin{document}
\title{Faraday Rotation Effect of Intracluster Magnetic Field on Cosmic
Microwave Background Polarization} \author{
Masahiro Takada\altaffilmark{1,3}, 
Hiroshi Ohno\altaffilmark{2}, and
Naoshi Sugiyama\altaffilmark{1}}

\altaffiltext{1}{
Division of Theoretical Astrophysics,
National Astronomical Observatory,
2-21-1 Osawa, Mitaka, Tokyo 181-8588, Japan 
}
\altaffiltext{2}{Research Center for the Early Universe, 
School of Science, University of Tokyo, Tokyo 113-0033, Japan}
\altaffiltext{3}{Department of Physics and Astronomy, 
University of Pennsylvania, 209 S. 33rd Street, 
Philadelphia, PA 19104, USA; mtakada@hep.upenn.edu}
\affil{%
mtakada@th.nao.ac.jp; 
ono@resceu.s.u-tokyo.ac.jp; naoshi@th.nao.ac.jp}

\begin{abstract}
The observed magnetic field of microgauss strength in clusters of
galaxies should induce the Faraday rotation effect on the linearly
polarized cosmic microwave background (CMB) radiation when the CMB
radiation propagates through a cluster at low redshift.  The Faraday
rotation arises from combined contributions of the magnetic field
strength, the electron density, the cluster size, and the
characteristic scale of the magnetic field along the line of sight.
Employing the Press-Schechter prescription for the cluster abundance
under the cold dark matter (CDM) scenario and a plausible isothermal
$\beta$-model for the gas distribution, we compute angular power spectra
of the CMB polarization fields including the Faraday rotation mixing
effect under the simple assumption of uniform magnetic field
configuration across a cluster.  As a result, we find that a {\em
parity-odd} $B$-type polarization pattern is statistically generated on
the observed sky, even when the primary polarization only contains the
{\em parity-even} $E$-mode component, such as in the case of pure scalar
perturbations. The generated $B$-type polarization has a peak with
amplitude of $\sim0.1~\mu{\rm K}(B_0/0.1~\mu{\rm G})(\nu_0/10~{\rm
GHz})^{-2}$ at angular scales of $l\approx 1000$ for the currently
favored adiabatic $\Lambda$CDM model.  This result also implies that, if
the magnetic field has a $0.5~\mu{\rm G}$ strength and we observe at
lower frequencies such as $\nu_0\simlt 5~{\rm GHz}$, the secondary
signal due to the Faraday rotation effect could be comparable to the
magnitude of the primary polarization.  The frequency dependence of the
Faraday rotation can be then used to discriminate the effect from
primary and other secondary signals on the CMB polarizations. Our
results therefore offer a new empirical opportunity to measure or
constrain the intracluster magnetic field in the average sense, combined
with measurements of the intracluster gas distribution through the X-ray
and SZ data.
\end{abstract}
%
\section{Introduction}
\label{intro} Various observations have revealed that clusters of
galaxies are pervaded by the strong magnetic field of microgauss
strength (e.g., see \cite{Kronberg94}; \cite{Carilli} for reviews and
references therein).  The multi-frequency Faraday-rotation measurements
of polarized radio sources inside or behind a cluster have been used to
estimate the magnetic field strength, combined with the X-ray data
(\cite{Vallee}; \cite{Kim}). Recently, Clarke, Kronberg, \& B\"ohringer
(2001) have drawn a firm conclusion that an intracluster hot plasma is
universally magnetized by $1-10\mu$G fields under the assumption of
$10-100$kpc field coherent scales, using $16$ normal low-$z$ {\it ROSAT}
cluster sample selected to be free of unusual strong radio halos,
widespread cooling flow and recent merger activity. However, except for
a few cases such as some background rotation measures per square degree
in the Coma cluster (\cite{Kim90}), it is generally difficult to measure
the angular profile of the magnetic field strength inside a cluster
because of the lack of the number of radio sources available per
cluster. The observations of cluster-wide diffuse radio halos have also
led to the evidence of the microgauss field, which is believed to be
synchrotron radiation by relativistic electrons accelerated in the shock
wave.  Moreover, in this case hard X-ray emission could be produced by
the same relativistic electron population through the inverse Compton
scattering of cosmic microwave background (CMB) photons. The combined
observation of hard X-ray emission and radio halos is therefore very
attractive in the sense that it allows us to directly estimate the
magnetic field strength without further restrictive assumptions for the
coherent length and the electron distribution. The recent detections of
hard X-ray emission have led to an independent estimation of
$\sim0.1\mu$G fields even in the outer ($\simgt 1{\rm~Mpc}$) envelopes
of clusters
(\cite{Bagchi}; \cite{Rephaeli99}; \cite{Fusco}).

However, the origin of the intracluster magnetic fields still rests 
a mystery in cosmology.  The following two scenarios have been
usually investigated in the literature. One is based on the idea that
the cluster field is related to the fields generated by dynamo mechanism
in individual galaxies and subsequent wind-like activity transports and
redistributes the magnetic fields in the intracluster medium
(e.g. \cite{Kronberg99} and references therein).
The general prediction of galactic dynamo mechanism is that the galactic
magnetic field could arise from an exponential amplification of a small
{\em seed} field during a galactic lifetime and energy of the mean
magnetic field could grow up to equipartition level with the turbulent
energy of fluid (\cite{Parker}; \cite{Zeld}; \cite{Chiba}).  It is not
still clear, however, that the galactic dynamo theory can explain the
detection of microgauss magnetic fields in high-$z$ damped Ly$\alpha$
absorption systems, which are supposed to be protogalactic clouds
(\cite{Welter}; \cite{Wolfe}).  On the other hand, an alternative scenario
is that the intracluster magnetic fields may grow via adiabatic
compression of a primordial field frozen with motions of cosmic plasma,
 where the primordial field is
assumed to be produced somehow during the initial stages of cosmic
evolution (e.g., \cite{Rees}; \cite{Kronberg94}; and also see
\cite{Grasso} for a recent review). Recently, using the cosmological,
magneto-hydrodynamic simulations of galaxy clusters, Dolag, Bartelmann,
\& Lesch (1999) have quantitatively shown that, if starting with the
initial magnetic field of $\sim10^{-9}$G strength, the final
intracluster field can be amplified by the gravitationally induced
collapsing motions to the observed strength of $\sim\mu$G irrespective
of uniform or chaotic initial field configurations motivated by
scenarios of the primordial or galactic-wind induced initial seed
fields on Mpc scales, respectively.
It is undoubtedly clear that the knowledge of the intracluster magnetic
field will lead to a more complete understanding of the physical
conditions of the intracluster medium and of the possible dynamical role
of magnetic fields (e.g., \cite{Loeb}; \cite{Vikhlinin}). Furthermore,
the magnetic field should play an essential role to the nonthermal
processes such as the synchrotron and high energy radiations and
possibly the cosmic ray production (\cite{LoebNat}; \cite{Totani};
\cite{Waxman}).  Hence, it is strongly desirable to perform further
theoretical and observational investigations on these issues in more
detail.

In this paper we investigate how the Faraday rotation effect due to the
intracluster magnetic field causes a secondary effect on the CMB
polarization fields. While Thomson scattering of temperature
anisotropies on the last scattering surface generates linearly polarized
radiation at the decoupling epoch (\cite{Kosowsky}; \cite{Kami},b;
\cite{ZaldPol}; \cite{HuWhite},b), the plane of linear polarization
should be rotated to some extent when the CMB radiation propagates
through the magnetized intracluster medium at a low redshift.  The great
advantage is that, since the CMB polarization field is a continuously
varying field on the sky, we can measure an angular profile of the
rotation measure in a cluster in principle, which cannot be achieved by
any other means. Furthermore, thanks to the frequency dependence of the
Faraday rotation, one will be able to separate this effect from the
primary signal and other secondary signals such as that induced by the
gravitational lensing effect due to the large-scale structure
(\cite{ZS98}).  In this paper, based on the Press-Schechter theory for
the cluster abundance under the cold dark matter (CDM) structure
formation scenario and a plausible isothermal $\beta$-model of the
intracluster gas distribution, we compute angular power spectra of the
CMB polarization fields including the Faraday rotation mixing effect
caused by clusters at low redshifts. As for the unknown field
configuration, we assume the uniform field of $\sim 0.1\mu{\rm G}$
strength across a cluster that is consistent with the recent Faraday
rotation measurements (Clarke et al. 2001).  This model allows us to
estimate the impact of this effect in the simplest way.
Our study thus proposes a new empirical opportunity to
measure or constrain the intracluster magnetic field in the average
sense that properties of the magnetic field could be extracted through
changes of the statistical quantities, CMB angular power spectra.

So far previous works have focused mainly on investigations of effects
of the primordial magnetic fields on the CMB temperature and
polarization anisotropies (\cite{KosoLoeb}; \cite{Adams}; \cite{Scan};
\cite{Harari}; \cite{Seshadri}; \cite{Mack}; and see also \cite{Grasso}
for a review).  Both the amplitudes of the cosmological primordial
magnetic field and the mean baryonic density rapidly increase with
redshift scaled as $(1+z)^2$ and $(1+z)^3$, respectively.  The dominant
contribution to the Faraday rotation effect is therefore imprinted
before decoupling ($z\simgt 10^3$), unless the structure formation at
low redshifts causes a significant amplification of the magnetic
field. Kosowsky, \& Loeb (1996) found that, if the current value of the
primordial field is of the order of $10^{-9}$G on Mpc scales
corresponding to the microgauss galactic field under the adiabatic
compression, the effect on the CMB polarization is potentially
measurable by satellite missions {\em MAP} and {\em Planck Surveyor}
(see also \cite{Scan}; \cite{Harari}).  It is also shown that such a
stochastic magnetic field in the early universe affects the temperature
fluctuations through the induced metric vector perturbations and thus
current measurements can put an upper limit of $\simlt 10^{-11}$G on the
current field amplitude (\cite{Mack}).

This paper is organized as follows. In Section \ref{RM} we first
construct a model to describe the magnetized hot plasma in a cluster by
assuming the uniform magnetic field configuration for simplicity.  We
then derive the angular power spectrum of the Faraday rotation angle
based on the Press-Schechter description of the cluster formation and
the $\beta$-model of the gas distribution.  In Section \ref{Formalism},
we present a formalism for calculating the angular power spectra of the
CMB polarization fields including the Faraday rotation mixing
effect. Section \ref{results} presents the results for the currently
favored CDM models. In Section \ref{Disc} we briefly present the
discussions and summary.  Unless stated explicitly, we assume the
favored $\Lambda$CDM cosmological model with $\Omega_{\rm m0}=0.3$,
$\Omega_{\lambda0}=0.7$, $\Omega_{b0}=0.05$, $h=0.7$, and $\sigma_8=1.0$
as supported from observations of CMB anisotropies and large-scale
structure (e.g. \cite{boom2}), where $\Omega_{m0}$, $\Omega_{b0}$, and
$\Omega_{\lambda0}$ are the present-day density parameters of
non-relativistic matter, baryon and the cosmological constant,
respectively, $h$ is the Hubble parameter and $\sigma_8$ denotes the rms
mass fluctuations of a sphere of $8h^{-1}$Mpc radius.  We also use the
$c=1$ unit for the speed of light.

\section{Model of Faraday Rotation due a Cluster Magnetic Field}
\label{RM}

\subsection{Faraday rotation of a magnetic field}

If monochromatic radiation of frequency $\nu$ is passing through a
plasma in the presence of a magnetic field $\bm{B}$ along the
propagation direction $\bm{\gamma}$, its linear polarization vector will
be rotated through the angle $\Delta\varphi$ (e.g., \cite{Rybicki}):
\begin{eqnarray}
\Delta\varphi&=&\frac{e^3}{2\pi m_e^2c^2\nu^2}
\int\!\!dl n_e(\bm{B}\cdot \bm{\gamma})\nonumber\\
&\approx &
8.12\times 10^{-2} (1+z)^{-2}\left(\frac{\lambda_0}{1{\rm ~cm}}\right)^{2}
\int\!\!\frac{dl}{{\rm ~kpc}}
\left(\frac{n_e}{1{\rm ~cm}^{-3}}\right)\left(\frac{\bm{B}\cdot\bm{\gamma}}
{1~\mu{\rm G}}\right),
\label{eqn:rm}
\end{eqnarray}
where $e$ and $m_e$ denote the charge and mass of electron,
respectively, $n_e$ is the number density field of electron along the
line of sight, and $\lambda_0$ is the observed wavelength.  Note that
the factor $(1+z)^{-2}$ comes from the relation of
$\lambda=\lambda_0(1+z)^{-1}$ and the sign of $\Delta\varphi$ could be
negative depending on the orientation of the magnetic field and the
reversal scale of magnetic field.  The equation above clearly shows that
the magnitude of the rotation measure depends on the electron number
density, the magnetic field strength and the characteristic scale length
of the field, and is larger for longer observed wavelengths because of
the dependence $\lambda_0^{2}$.  Thus, as a possible source of the
Faraday rotation, in this paper we consider the magnetized hot plasma in
clusters of galaxies, motivated by the fact that the deep gravitational
potential well of dark matter makes the intracluster gas dense and
highly ionized and many observations have revealed the existence of the
strong magnetic field with $\sim ~\mu{\rm G}$ strength(e.g., Clarke et
al. 2001).

\subsection{Cluster model with a uniform magnetic field}
To calculate the Faraday rotation measure for the magnetized
intracluster hot plasma, we have to assume how the magnetic field is
distributed in a cluster along the line of sight.  There are two
possibilities considered; the uniform field configuration and, more
realistically, the tangled field, where the former field has constant
strength and direction through the entire cluster while the latter field
has reversal scales along the line of sight. For the Coma cluster, Kim
et al. (1990) suggested $\sim 10{\rm ~kpc}$ for the field reversal scale
from the observed rotation measures of some sources close in angular
position to the cluster, and Feretti et al. (1995) discovered the
magnetic field structure with smaller scales down to $\sim 1{\rm ~kpc}$
from the rotation measures of radio halo in the Coma, leading to the
stronger field of $\sim 10\mu{\rm G}$ on the scale. On the other hand,
there is observational evidence of coherence of the rotation measures
across large radio sources, which indicates that there is at least a
field component smoothly distributed across cluster scales of $\sim
100{\rm ~kpc}$ (e.g., \cite{Taylor}). Thus, the magnetic field structure
is still unknown because of the lack of the number of sources available
per cluster for measuring the magnetic field. The angular profile of the
rotation measure compiled from all sample of radio sources for 16
different normal clusters (Clarke et al. 2001) yields the field
strengths between $\sim0.5$ and $\sim 3~\mu{\rm G}$ for a uniform field
model, while $\sim 5~\mu{\rm G}(l/10{\rm ~kpc})^{-1/2}(h/0.75)^{1/2}$
for a simple tangled-cell model with a constant coherence length $l$. In
this paper, we consider a uniform magnetic field model for the sake of
revealing the impact of the Faraday rotation effect on the CMB
polarization in the simplest way.
We then assume that the amplitude of magnetic field 
at the cluster formation epoch is universally constant, namely it does not 
depend on redshift; $B(z_{\rm form})=B_0={\rm constant}$. 

Next let us consider a model to describe the gas distribution in a
cluster.  Motivated by the X-ray observations, we adopt the spherical
isothermal $\beta$-model as for the gas density profile of each cluster
expressed by
\begin{equation}
\rho(r)=\rho_0\left[1
+(r/r_c)^2\right]^{-3\beta/2},
\label{eqn:beta}
\end{equation}
where $\rho_0$ is the central gas mass density and $r_c$ is the core
radius. The observed values of $\beta$, obtained from X-ray surface
brightness profiles, range from $0.5$ to $0.7$ (e.g. see \cite{Jones}).
In the following discussion we assume $\beta=2/3$ for simplicity.  Note
that Makino, Sasaki and Suto (1997) showed that the universal density
profile of dark matter halo (\cite{NFW}) can reproduce the $\beta$-model
for massive clusters under assumptions of the hydrostatic equilibrium
and isothermality. In the spherical collapse model, the virial radius
$r_{\rm vir}$ for the halo with mass $M$ is obtained from
\begin{equation}
M=\frac{4\pi}{3}\rho_{\rm cr}(z)\Delta(z)r_{\rm vir}^3,
\end{equation}
where $\Delta(z)$ is the overdensity of collapse and the fitting formula
is given by Bryan \& Norman (1998), and $\rho_{\rm cr}(z)$ is the
critical matter density at redshift $z$ defined by $\rho_{\rm
cr}(z)=3H_0^2/(8\pi G)(1+z)^3$.

Once the uniform magnetic field and the gas density profile are given,
we can easily calculate the angular profile of the Faraday rotation angle
for a cluster projected on the sky similarly as the procedure to
calculate the profile of the thermal Sunyaev-Zel'dovich (SZ) effect
(\cite{Atrio}; \cite{KK}). Using the gas density profile
(\ref{eqn:beta}) and equation (\ref{eqn:rm}), the angular profile can be
calculated as
\begin{eqnarray}
\Delta\varphi(\theta,z)&=&
\frac{e^3 n_{\rm e0}\rcor}{2\pi m_e^2c^2\nu_0^2(1+z)^2}
(B\cos\chi)\frac{2}{\sqrt{1+(\theta/\theta_c)^2}}
\tan^{-1}\sqrt{\frac{p^2-(\theta/\theta_c)^2}{1+(\theta/\theta_c)^2}},
\label{eqn:angrm}
\end{eqnarray}
where $p\equiv r_{\rm vir}/r_c$, $\chi$ is the relative angle between
the direction of the magnetic field and the line of sight, $n_{\rm e0}$ is
the central number density of electron, and $\theta$ is the angular
separation between the line of sight and the cluster center. The
angular core size $\theta_c$ is related to $r_c$ via
$\theta_c=\rcor/d_A(z)$, where $d_A(z)$ is the angular diameter distance
to the cluster at redshift $z$. Note that $p$, given as the ratio between
the viral and the core radius, is not a parameter but a function of mass
$M$ and redshift $z$ in a general case.

\begin{figure}[t]
 \begin{center}
     \leavevmode\epsfxsize=9cm \epsfbox{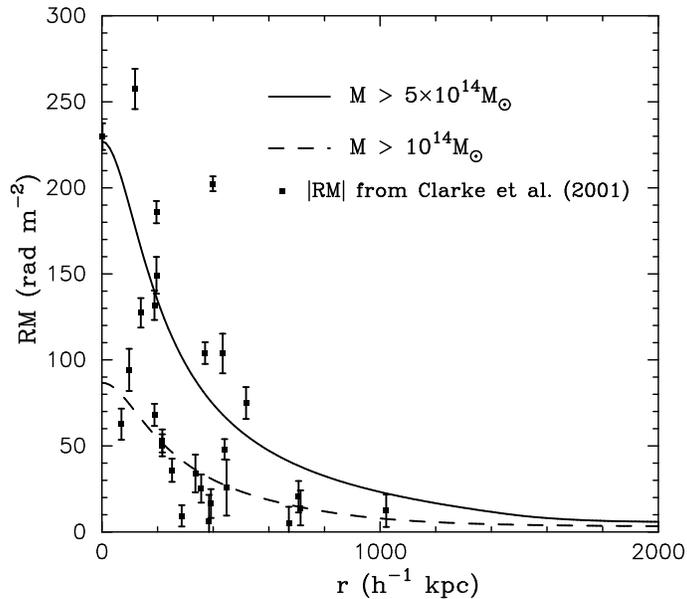}
\caption{Angular profile of rms of the rotation measure for a cluster with
$M\ge 5\times 10^{14}M_\odot$ as a function of the radius from the
cluster center (solid line), which is computed using equation
(\ref{eqn:averm}).  We hare considered $B_0=0.2\mu{~\rm G}$ for the
intracluster magnetic field strength. In order to demonstrate the
dependence of the minimum mass used in the integration of equation
(\ref{eqn:averm}), the result for $M\ge 10^{14}M_{\odot}$ is also shown
by dashed line.  The mass range of $M\ge 5\times 10^{14}M_\odot$
corresponds to the $X$-ray luminosity of $L_X\simgt 10^{44}{\rm
erg~s}^{-1}$ in the $0.1-2.4{\rm ~keV}$ energy band, which roughly
corresponds to the {\em ROSAT} sample of the Faraday rotation
measurements used by Clarke et al. (2001).  For comparison, square
symbols with error bars represent the observational data from Clarke et
al. (2001), where the error arises from the Faraday rotation
contribution due to our Galaxy.  \label{fig:averm}}
 \end{center}
\end{figure}

We also have to relate the quantities $\rcor$ and $n_{e0}$ to the total
mass $M$ and redshift $z$.  Assuming the full ionization for
intracluster gas, the central electron number density is given by
$n_{e0}= (\rho_{\rm gas0}/m_p)(1-Y/2)$, where $Y$ is the helium mass
fraction and we fix $Y=0.24$ in this paper.  It is known that the
formation of cluster core regions contains large uncertainties.  We here
employ the {\it entropy driven} model (\cite{Kaiser91}; \cite{Evrard})
to describe the core radius, motivated by the fact that this model can
explain observations such as the X-ray luminosity function better than
the simplest self-similar model (\cite{Kaiser86}) does.  In this model
the core is supposed to be in a minimum entropy phase $s_{\rm
min}=c_v\ln(T_{e0}/\rho_{\rm gas0}^{\gamma-1})$, where $T_{e0}$ denotes
the central electron temperature, $c_v$ is the specific heat capacity of
the gas at constant volume, and $\gamma$ is the ratio of specific heats
at constant pressure and constant volume.  If using $M=(4\pi/3)\rho_{\rm
cr}(z)\Delta(z)r_{\rm vir}^3 \approx 4\pi\rcor^2\rvir(\rho_{\rm
gas0}/f_{\rm gas})$ for $\rvir/\rcor \gg 1$, we can obtain
\begin{eqnarray}
\frac{\rvir}{\rcor}(M,z)&\approx& \left[\frac{
3\rho_{\rm gas0}(M,z)/f_{\rm gas}}{\rho_{\rm cr}(z)\Delta_c(z)}
\right]^{1/2}\nonumber\\
&\propto& \frac{T_{e0}^{1/[2(\gamma-1)]}(M,z)}{\Delta_c^{1/2}(z)}
(1+z)^{-3/2},
\label{eqn:rcore}
\end{eqnarray}
where $f_{\rm gas}$ is the gas mass fraction for a cluster and we assume
that $f_{\rm gas}$ is equal to the cosmological mean; $f_{\rm gas}\equiv
M_{\rm gas}/M=\Omega_{\rm b0}/\Omega_{m0}$ (\cite{White93}). The adiabatic
index is fixed at $\gamma=5/3$ throughout this paper.  Moreover, we
assume the isothermality and that the central electron temperature
$T_{e0}$ is equal to the virial temperature $T_{\rm vir}$ given by
\begin{equation}
k_BT_{\rm vir}=\eta\frac{\mu m_p GM}{3\rvir},
\end{equation}
with $\mu=0.59$. The quantity $\eta$ is a fudge factor of order unity
that should be determined by the efficiency of the shock heating of the
gas (\cite{Eke98}), and we fix $\eta=3/2$.  If the proportional constant
to calculate $\rcor$ in equation (\ref{eqn:rcore}) is determined by
assuming $\rcor(10^{15}h^{-1}\solam,z=0)=0.15h^{-1}{\rm Mpc}$, we can
obtain the core radius for an arbitrary cluster with mass $M$ and at
redshift $z$. This cluster model then yields $\rvir/\rcor=13.6$, $n_{\rm
e0}=5.0\times 10^{-3}{\rm ~cm}^{-3}$, and $T_{\rm e}=6.5 {\rm keV}$ for
$M=10^{15}h^{-1}\solam$ at $z=0$ in the $\Lambda$CDM model. Using these
typical values of physical quantities for a cluster with $M=
10^{15}h^{-1}\solam$, equation (\ref{eqn:angrm}) tells us the magnitude
of the rotation measure at the cluster center as
\begin{equation}
\Delta\varphi(z=0,\theta=0)\approx 2.34\times 10^{-1}{\rm ~rad}
\left(\frac{\nu_0}{10{\rm ~GHz}}
\right)^{-2}\left(\frac{B_0\cos\chi}{0.1 \ \mu{\rm G}}\right).
\label{eqn:orderest}
\end{equation}
Thus, if the line-of-sight component of the uniform magnetic field has a
strength of $\sim 0.1~\mu{\rm G}$ and we can observe at low frequencies
such as $10~{\rm GHz}$, the magnitude of the rotation measure angle
could be of the order of $10^{-1}{\rm rad}~(\approx 6^\circ)$ for such a
massive cluster.  The observed scatter of the normalized rotation
measure (${\rm RM}\equiv \Delta\varphi/\lambda_0^2[{\rm m}]$) is ${\rm
RM} =0-250{\rm ~rad~m}^{-2}$ at the radius in the range of $0\simlt
r\simlt 1h^{-1}{\rm ~Mpc}$ from the cluster center (Kim et al. 1991;
Clarke et al. 2001).  The value estimated in equation
(\ref{eqn:orderest}) corresponds to ${\rm RM}\approx 260 {\rm ~rad\
m}^{-2}$ at the cluster center for the typical case of $B_0\cos\chi=0.1\
\mu{\rm G}$.  Our model thus seems consistent with the
observations. This can be more quantitatively verified as follows. Using
our model of the rotation measure angle (\ref{eqn:angrm}) and the
Press-Schechter mass function $dn/dM$ (\cite{PS}) (see equation
(\ref{eqn:PS})), we can compute the variance of the rotation measure for
present-day clusters against the radius from the cluster center:
\begin{equation}
\skaco{\rm RM^2}(r)=\left.\int^{M_{\rm max}}_{M_{\rm min}}\!\!dM\frac{dn}{dM}
\left[\frac{\Delta\varphi(r,z=0)}{\lambda^2}\right]^2\right/
\int^{M_{\rm max}}_{M_{\rm min}}\!\!dM\frac{dn}{dM}.
\label{eqn:averm}
\end{equation}
In the calculation of the equation above, we use
$\skaco{\cos^2\chi}=1/3$ for the ensemble average of angle $\chi$
between the magnetic field direction and the line of sight, since the
field orientation is considered to be random for each cluster. The solid
line in Figure \ref{fig:averm} shows the average profile
$\sqrt{\skaco{{\rm RM}^2}(r)}$ for $B_0=0.2\mu{\rm G}$ in the mass range
of $M\ge 5\times 10^{14}M_\odot$.  Note that this mass range corresponds
to $L_X\simgt 10^{44} {\rm erg~s}^{-1}$ for the $X$-ray luminosity in
the ROSAT $0.1-2.4~{\rm keV}$ band, where only the thermal
bremsstrahlung emission is considered. Thus, this $X$-ray luminosity
range is roughly consistent with the ROSAT sample of the Faraday
rotation measurements used by Clarke et al. (2001).  For comparison,
square symbols represent the observational data from Clarke et
al. (2001) with $\pm 1\sigma$ errors arising from the Faraday
contributions due to our Galaxy.  It is clear that our model with
$B_0=0.2\mu{\rm G}$ can roughly reproduce the observed magnitude and
dispersion of the rotation measures as a function of the radius. In this
sense, it is likely that we can at least estimate a correct magnitude of
the Faraday rotation effect on the angular power spectra of CMB
polarization fields, even though the uniform magnetic field is an
unrealistic model, because the contribution to the angular power spectra
comes from the second moments of the rotation measure angle by
definition (see equation (\ref{eqn:rmclB})).  Furthermore, there could
be an uncertain selection effect in the observational data.  For
example, one cluster sample with the large RM value of $|{\rm RM}|=229$
at the radius of $r=3~{\rm kpc}$ has the $X$-ray luminosity of
$L_X=2.0\times 10^{42}~{\rm erg~s}^{-1}$ or the mass of $M\approx
7.0\times 10^{13}M_\odot$, which is smaller than the lowest mass of
$M=5\times 10^{14}M_\odot$ used for the theoretical prediction shown by
the solid line. To illustrate the sensitivity of the lowest mass to the
theoretical prediction, the dashed line in Figure \ref{fig:averm} shows
the result of equation (\ref{eqn:averm}) for $M\ge 10^{14}M_\odot$
($L_X\simgt 4.4\times 10^{42}~{\rm erg~s}^{-1}$) and $B_0=0.2\mu{\rm
G}$, and it clearly underestimates the observational result.  For the
similar reason, it is thought that Clarke et al. (2001) derived a
higher value of $\sim 0.5~\mu{\rm G}$ for the uniform magnetic field
model from their observational data.  In addition to these facts, there
is possible contamination of line emissions from the intracluster metal
to the observed $X$-ray luminosity.  Therefore, our model of
$B_0=0.2~{\mu G}$ often used in the following discussion could lead to
the minimum magnitude of the Faraday rotation effect on the CMB
polarization power spectra for a realistic effect.
%

As will be shown in detail, the Faraday rotation effect generates
secondary patterns of the CMB polarizations expressed as $Q\sim
\Delta\varphi \tilde{U}$ and $U\sim \Delta\varphi \tilde{Q}$ in terms of
the Stokes parameters $Q$ and $U$, where $\tilde{Q}$ and $\tilde{U}$ are
those parameters of the primary polarization generated via Thomson
scattering at the decoupling ($z_\ast\approx 10^3$). Hence, equation
(\ref{eqn:orderest}) implies that, if we observe the CMB polarization
through a cluster at longer wavelengths such as $\lambda_0\simgt
6{\rm~cm}$ ($\nu_0\simlt 5{\rm ~GHz}$), the secondary signals could be
comparable to the primary signals for massive clusters with
$B_0\cos\chi=0.1\mu{\rm G}$.  In this case,
we could directly measure the Faraday rotation effect in an individual
cluster like the SZ effect on the CMB temperature fluctuations. This
would be also true for particular clusters that have an unusual stronger
magnetic field such as $\simgt 1~\mu{\rm G}$.  Moreover, since the
primary CMB polarization field has very smooth structure on angular
scales of $\theta\simlt 10'$ because of the Silk damping (\cite{Silk};
and also see \cite{HuSugi} in more detail), the angular profile of
the rotation measure could be directly reconstructed in principle from the
secondary signals generated on small scales even for a single observed
frequency.  These interesting possibilities are now being investigated
in detail, and the results will be presented elsewhere (Ohno et al. 
2001 in preparation).

\subsection{Angular power spectrum of the rotation measure}
We are interested in the secondary effect on the angular power spectra
of the CMB polarization due to the Faraday rotation effect caused by
clusters. We thus need to compute the angular power spectrum of the
rotation measure angle $\Delta\varphi(\theta)$, which gives the strength
of $\Delta\varphi$ in the average sense for the all-sky survey at
two-point statistics level.  The angular Fourier transformation of the
rotation measure (\ref{eqn:angrm}) is given by
\begin{equation}
\Delta\varphi_{l}(z)=2\pi\int^{\theta_{\rm vir}}_0\!\!
d\theta \theta
\Delta\varphi(\theta,z)J_0(l\theta). 
\end{equation}
where $\theta_{\rm vir}\equiv r_{\rm vir}/d_A(z)$, $J_0(x)$ is the
zero-th order Bessel function, and we have used the small-angle
approximation (Bond \& Efstathiou 1987).  Since clusters are discrete
sources, contributions to the power spectrum of the rotation measure are
divided into the Poissonian contribution, $C^{(P)}_l$, and the
contribution from the spatial correlation between clusters, $C^{(C)}_l$
(Peebles 1980).  From equation (\ref{eqn:angrm}) the correlation
contribution $C^{(C)}_l$ depends on the ensemble average
$\skaco{\cos\chi_1\cos\chi_2}$, where $\chi_1$ and $\chi_2$ are the
relative angles between the magnetic fields and the line of sights for
those two clusters, respectively, while $C^{(P)}_l$ depends on
$\skaco{\cos^2\chi }$, leading to the non-vanishing contribution for any
distribution of $\chi$. If the origin of the magnetic field is
primordial, the correlation between two intracluster magnetic fields
could be non-vanishing even on separation scales between clusters like
that between the density fluctuation fields, and in this case we have
some contribution from $C^{(C)}_l$ because
$\skaco{\cos\chi_1\cos\chi_2}\ne0$.  However, it is known that the
Poissonian contribution always dominates the correlation one because
clusters are rare objects (\cite{KK}). For this reason, we here assume
$\skaco{\cos\chi_1\cos\chi_2}=0$ and consider only the Poisson
contribution $C^{(P)}_l$ for simplicity.  Following the method developed
by Cole \& Kaiser (1988) and Limber's equation, we can obtain an
integral expression of $C^{(C)}_l$;
\begin{eqnarray}
C_l^{\rm RM(P)}&=&\int^{z_{\rm dec}}_0\!\!dz\frac{dV}{dz}
\int^{M_{\rm max}}_{M_{\rm min}}\!\!dM\frac{dn(M,z)}{dM}
|\Delta\varphi_{l}(M,z)|^2,\label{eqn:clpoiss}
\end{eqnarray}
where $V(z)$ and $r(z)$ are the comoving volume and the comoving
distance, respectively, and $dn/M(M,z)$ is the comoving number density
for halos with mass in the range of $M$ and $M+dM$ at redshift $z$. We
compute $dn/dM$ using the Press-Schechter model (\cite{PS}), and in this
case we have
\begin{equation}
\frac{dn(M,z)}{dM}=\sqrt{\frac{2}{\pi}}\frac{\bar{\rho}}{M}
\left|\frac{d\ln \sigma}{dM}\right|\nu_c e^{-\nu_c^2/2},
\label{eqn:PS}
\end{equation}
where $\nu_c=\delta_c(z)/\sigma(M)$ denotes the threshold, $\sigma(M)$
is the present-day rms mass fluctuation within the top-hat filter
corresponding to the mass scale of $M$, $\delta_c(z)\approx 1.69/D(z)$
is the threshold overdensity of the spherical collapse model, and $D(z)$ is
the linear growth factor of density fluctuations.  

\begin{figure}[t]
 \begin{center}
     \leavevmode\epsfxsize=10cm \epsfbox{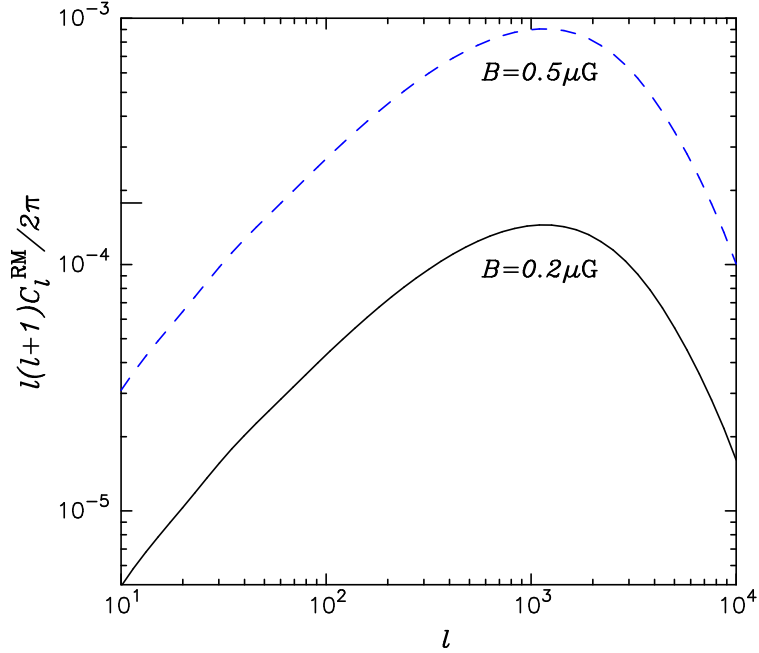}
\caption{The angular power spectra of the rotation measure angle due the
intracluster magnetic field calculated by equation (\ref{eqn:clpoiss})
for the observed frequency of $10{\rm GHz}$ in the LCDM model
($\Omega_{m0}=0.3$, $\Omega_{\lambda0}=0.7$, $h=0.7$,
$\Omega_{b0}=0.05$, $\sigma_8=1$). We here considered two cases of the
uniform magnetic fields $B_0$ of $0.2~\mu{\rm G}$ (solid line) and
$0.5~\mu{\rm G}$ (dashed) strength across a cluster.  Note that the
power spectrum has the dependence on the magnetic field strength and the
observed frequency as $\propto B_0^2\nu_0^{-4}$.  \label{fig:clrm}}
 \end{center}
\end{figure}

Figure \ref{fig:clrm} shows the angular power spectra of the rotation
measure through clusters for $B_0=0.2~\mu{\rm G}$ (solid
line) and $0.5~\mu{\rm G}$ (dashed) assuming the observed frequency of
$\nu_0=10{\rm ~GHz}$ in the $\Lambda$CDM model.  Note that we have again
used $\skaco{\cos^2\chi}=1/3$.
We have confirmed that dominant contribution to the power spectrum comes
from the massive clusters with mass larger than $10^{14}M_\odot$ at low
redshifts such as $z\simlt 0.5$, similarly as the results of SZ effect
due to clusters (e.g., \cite{Cooray}).

\section{Secondary Effect of Faraday Rotation in a Cluster 
on CMB Polarization}
\label{Formalism}

In this section we develop a formalism for calculating the secondary
Faraday rotation effect induced by magnetized intracluster hot plasma on
the CMB polarization fields and their angular power spectra. 

The properties of CMB polarization fields can be described in terms of
the Stokes parameters $Q$ and $U$ (\cite{Kosowsky}; \cite{Kami},b;
\cite{ZaldPol}).  Let us consider quasi-monochromatic waves propagating
in the z-direction, in which the electric and magnetic fields vibrate on
the $x$-$y$ plane. We can then express the electric field $\bm{E}$ in
terms of the complex orthogonal unit vectors of circularly
polarized waves as $\bm{E}=(E_{\rm R}e^{i\delta_{\rm
R}}\bm{\epsilon}_R+E_{\rm L}e^{i\delta_{\rm L}}\bm{\epsilon}_{\rm
L})e^{-i\omega t}$, where $\bm{\epsilon}_{\rm R,L}\equiv
(\bm{\epsilon}_x \mp i\bm{\epsilon}_y)/\sqrt{2}$ and $\bm{\epsilon}_{\rm
R}$ and $\bm{\epsilon}_{\rm L}$ are the unit vectors of right-hand and
left-hand polarized waves, respectively. Using those quantities, the
Stokes parameters $Q$ and $U$ measured in the fixed $(x,y)$-coordinates
can be expressed as
\begin{eqnarray}
Q&\equiv&|\bm{E}\cdot\bm{\epsilon}_x|^2-|\bm{E}\cdot\bm{\epsilon}_y|^2
=2E_{\rm R}E_{\rm L}\cos(\delta_{\rm R}-\delta_{\rm L}),\nonumber\\
U&\equiv&2{\rm Re}\left[(\bm{E}\cdot\bm{\epsilon}_x)^{\ast}
(\bm{E}\cdot\bm{\epsilon}_y)\right]=2E_{\rm R}E_{\rm L}
\sin(\delta_{\rm R}-\delta_{\rm L}). 
\label{eqn:QUparas}
\end{eqnarray}
If there is an intervening magnetic field in the ionized plasma between
the last scattering surface and the present epoch, the right-hand and
left-hand polarized waves travel with different phase velocities by the
Faraday rotation effect.  Therefore, from equation (\ref{eqn:QUparas}),
we can derive the following equation to govern the time evolution of $Q$
and $U$ in the presence of the magnetized plasma;
\begin{eqnarray}
\frac{dQ}{dt}&=-&U\frac{d(\delta_{\rm R}-\delta_{\rm L})}{dt}
=-2\omega_BU,\nonumber\\
\frac{dU}{dt}&=&2\omega_BQ,
\label{eqn:boltz}
\end{eqnarray}
where $2\omega_B\equiv d(\delta_{R}-\delta_{\rm L})/dt$, $t$ is the
cosmic time and we have ignored other effects associated with the
Thomson scattering. Note that $\omega_B$ is also related to the rotation
measure angle $\Delta\varphi$ defined by equation (\ref{eqn:rm}) as
$\omega_B=d(\Delta\varphi)/dt$. Equation (\ref{eqn:boltz}) clearly shows
that the Faraday rotation causes the mixing between $Q$- and $U$-type
polarizations. More importantly, even if the primary CMB anisotropies
generate only the $Q$-type polarization in the statistical sense
(\cite{ZaldPol}), the Faraday rotation mixing induces the $U$-type
polarization on the observed sky from the primary $Q$ mode.

Since we are interested in the secondary effect on the CMB polarizations,
the solutions of equation (\ref{eqn:boltz}) for $Q$ and $U$ along the line
of sight with direction of $\bm{\theta}$ can be approximately solved  up
to the first order of $\omega_B$ as
\begin{eqnarray}
Q(t_0,\bm{\theta})
&\approx&\tilde{Q}(t_\ast,\bm{\theta})-2\tilde{U}(t_\ast,\bm{\theta})
\int^{t_0}_{t_\ast}\!\!dt\omega_B(t,\bm{\theta})\nonumber\\ 
U(t_0,\bm{\theta})&\approx& \tilde{U}(t_\ast,\bm{\theta}) 
+2\tilde{Q}(t_\ast,\bm{\theta})\int^{t_0}_{t_\ast}
\!\!dt\omega_B(t,\bm{\theta}),
\label{eqn:magPol}
\end{eqnarray}
where $\tilde{U}$ and $\tilde{Q}$ denote the primary $U$- and
$Q$-polarization fluctuations generated at the decoupling epoch
($z_\ast\approx 10^3$), respectively, and we have neglected other
secondary polarization fields generated at low redshifts that have
insignificant power compared to the primary signal on relevant angular
scales for a realistic reionization history (e.g., \cite{Hu}). The
quantities $t_\ast$ and $t_0$ are the cosmic time at the decoupling and
the present time, respectively.  In what follows quantities with tilde
symbol denote quantities at $t_\ast$.
 
The Stokes parameters $Q$ and $U$ depend on the choice of coordinates
and are not invariant under the coordinate rotation. For this reason, it
is useful to introduce the new bases of rotationally invariant
polarization fields, called the parity-even electric mode ($E$) and the
parity-odd magnetic mode ($B$) (\cite{ZaldPol}; \cite{Kami},b;
\cite{HuWhite},b).  Any polarization pattern on the sky can be
decomposed into these modes.  Since the secondary effect due to clusters
of galaxies on the CMB polarization is important only on small angular
scales such as $\theta\simlt 1^\circ$, we can safely employ the
small-angle approximation (\cite{ZS98}).  Based on these considerations,
$Q$ and $U$ fields can be expressed using the two-dimensional Fourier
transformation in terms of the electric and magnetic modes as
\begin{eqnarray}
Q(\bm{\theta})&=&\int\!\!\frac{d^2\bm{l}}{(2\pi)^2}
\left[E_{\bm{l}}\cos2\phi_{\bm{l}}-
B_{\bm{l}}\sin2\phi_{\bm{l}}\right]e^{i\bm{l}\cdot\bm{\theta}},\nonumber\\
U(\bm{\theta})&=&\int\!\!\frac{d^2\bm{l}}{(2\pi)^2}
\left[E_{\bm{l}}\sin2\phi_{\bm{l}}+B_{\bm{l}}\cos2\phi_{\bm{l}}\right]e^{i\bm{l}
\cdot\bm{\theta}},
\end{eqnarray}
where $\phi_{\bm{l}}$ is defined by $l_x+il_y=l\exp(i\phi_{\bm{l}})$, and
$E_{\bm{l}}$ and $B_{\bm{l}}$ are the two-dimensional Fourier
coefficients for the $E$- and $B$-modes, respectively.
The Fourier components $\tilde{E}_{\bm{l}}$ and $\tilde{B}_{\bm{l}}$ 
for the primary CMB polarization fields satisfy
\begin{eqnarray}
\kaco{\tilde{E}_{\bm{l}}\tilde{E}^\ast_{\bm{l}'}}
=(2\pi)^2 C_{\tilde{E}l}\delta^2(\bm{l}-\bm{l}'),\nonumber\\
\kaco{\tilde{B}_{\bm{l}}\tilde{B}^\ast_{\bm{l}'}}
=(2\pi)^2 C_{\tilde{B}l}\delta^2(\bm{l}-\bm{l}'),
\end{eqnarray}
where $C_{\tilde{E}l}$ and $C_{\tilde{B}l}$ are the angular power
spectra of primary $E$- and $B$-modes, respectively.  The primary
$B$-mode polarization is generated by the quadrupole temperature
anisotropies associated with the tensor (gravitational wave) and vector
perturbations.  If we assume the standard inflation-motivated model with
a spectral index of $n_{\rm s}=1$ for the scalar perturbations, the
tensor and vector perturbations are sufficiently smaller than the scalar
perturbations. In the following discussion we ignore the primary
$B$-mode power spectrum for simplicity; $C_{\tilde{B}l}=0$.

From equation (\ref{eqn:magPol})
the two-point correlation functions of $Q$ and $U$ fields 
including the Faraday rotation mixing effect can be computed as
\begin{eqnarray}
C_Q(\theta)&\equiv&\kaco{Q(0)Q(\bm{\theta})}
\approx C_{\tilde{Q}}(\theta)+4C_{\tilde{U}}(\theta)
C^{{\rm RM}}(\theta),\nonumber\\
C_U(\theta)&\equiv&\kaco{U(0)U(\bm{\theta})}
\approx C_{\tilde{U}}(\theta)+4C_{\tilde{Q}}(\theta)
C^{{\rm RM}}(\theta),
\end{eqnarray}
with
\begin{eqnarray}
C_{\tilde{Q}}(\theta)&=&\int\!\!\frac{ldl}{2\pi}\frac{C_{\tilde{E}l}}{2}
(J_0(l\theta)+J_4(l\theta)),\nonumber\\
C_{\tilde{U}}(\theta)&=&\int\!\!\frac{ldl}{2\pi}\frac{C_{\tilde{E}l}}{2}
(J_0(l\theta)-J_4(l\theta)),
\label{eqn:magtwo}
\end{eqnarray}
where we have employed a reasonable assumption that the primary fields
($\tilde{Q}$ and $\tilde{U}$) on the last scattering surface and the
rotation measure field $\omega_B$ induced by low-$z$ clusters are
statistically uncorrelated.  Note that $C^{{\rm RM}}(\theta)$ is the
two-point correlation function of the rotation measure angle and can be
expressed in terms of the angular power spectrum of the rotation measure
as
\begin{equation}
C^{\rm RM}(\theta)\equiv
\kaco{\int\!\!dt \omega_B(t,0)\int\!\!dt'
\omega_{B}(t',\bm{\theta})}=
\int\!\!\frac{ldl}{2\pi}C^{\rm RM(P)}_lJ_0(l\theta),
\end{equation}
where $C^{\rm RM(P)}_l$ is given by equation (\ref{eqn:clpoiss}). 

We are now in the position to derive the angular power spectra of $E$
and $B$ modes in the observed CMB sky. Following the method developed by
Zaldarriaga, \& Seljak (1998), $C_{El}$ and $C_{Bl}$ can be expressed as
\begin{eqnarray}
C_{El}&=&\pi\int^\pi_0\!\theta d\theta \left[
(C_Q(\theta)+C_U(\theta))J_0(l\theta)+(C_Q(\theta)-C_U(\theta))J_4(l\theta)
\right],\nonumber\\
C_{Bl}&=&\pi\int^\pi_0\!\theta d\theta \left[
(C_Q(\theta)+C_U(\theta))J_0(l\theta)-(C_Q(\theta)-C_U(\theta))J_4(l\theta)
\right].
\end{eqnarray}
Therefore, from equation (\ref{eqn:magtwo}) we have
\begin{eqnarray}
C_{El}&=&C_{\tilde{E}l}+\int\!\!l'dl'W^E_{ll'}C_{\tilde{E}l'}\nonumber\\
C_{Bl}&=&\int\!\!l'dl'W^B_{ll'}C_{\tilde{E}l'},
\label{eqn:rmclB}
\end{eqnarray} 
where $W_{Ell'}$ and $W_{Bll'}$ are the window functions defined by
\begin{eqnarray}
W^E_{ll'}&\equiv& 2\int_0^\pi\!\theta d\theta
\left[J_0(l\theta)J_0(l'\theta)-J_4(l\theta)J_4(l'\theta)\right]
C^{\rm RM}(\theta),\nonumber\\
W^B_{ll'}&\equiv& 2\int_0^\pi\!\theta d\theta 
\left[J_0(l\theta)J_0(l'\theta)+J_4(l\theta)J_4(l'\theta)\right]
C^{\rm RM}(\theta).
\label{eqn:window}
\end{eqnarray}
Equations (\ref{eqn:rmclB}) clearly show that the Faraday rotation
mixing generates the non-vanishing angular power spectrum of the
$B$-mode polarization from the primary $E$-mode.

\section{Results and Estimations of the Detectability}
\label{results} Throughout this paper, we employ the $\Lambda$CDM
model with $\Omega_{m0}=0.3$, $\Omega_{\lambda0}=0.7$, $\Omega_{b0}=0.05$,
$h=0.7$ and $\sigma_8=1.0$ as supported from observations of the CMB
anisotropies and the large-scale structure.  As for the power spectrum
of dark matter used in the calculation of the Press-Schechter mass
function, we adopted the Harrison-Zel'dovich spectrum and the BBKS
transfer function (\cite{BBKS}) with the shape parameter from Sugiyama
(1995).

\begin{figure}[t]
 \begin{center}
     \leavevmode\epsfxsize=15cm \epsfbox{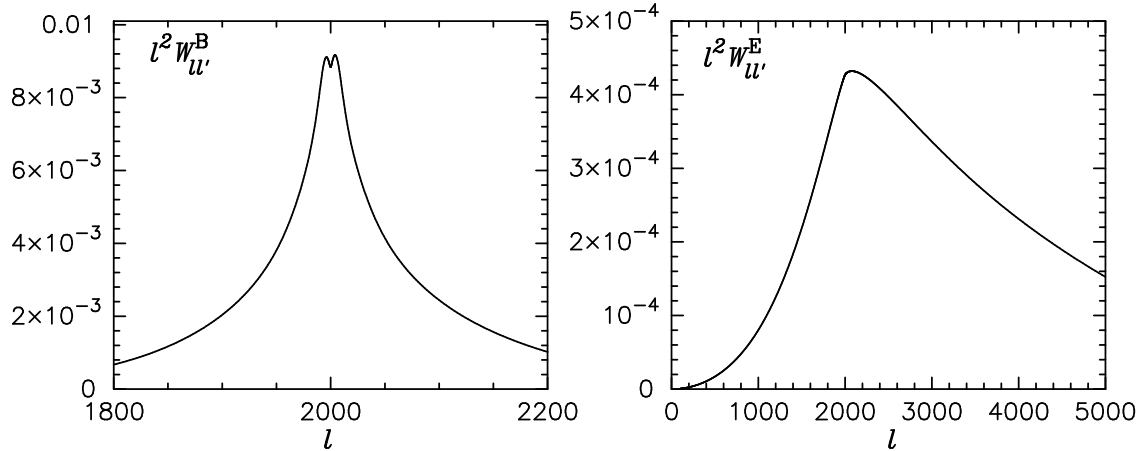}
\caption{This figure shows the weighted window functions $l^2W^B_{ll'}$
and $l^2W^E_{ll'}$ defined by equation (\ref{eqn:window}) as a function of
$l$ for a given mode $l'=2000$.  We here present their power per
logarithmic interval in $l$.  We used the model of the magnetic field
strength $0.2~\mu{\rm G}$ and the observed frequency $10~{\rm GHz}$.
 \label{fig:window}}
 \end{center}
\end{figure}

Figure \ref{fig:window} shows the window functions of $B$- and $E$-modes
per logarithmic interval in $l$, $l^2W^{B}_{ll'}$ and $l^2W^E_{ll'}$,
for a given mode $l'=2000$ as derived by equation (\ref{eqn:window}). We
consider the magnetic field strength $B_0=0.2~\mu{\rm G}$ and the observed
frequency $10~{\rm GHz}$ in the $\Lambda$CDM model.  One can see that
the window functions for the mixing between $E$ and $B$ modes are well
localized in $l$ space.

\begin{figure}[t]
 \begin{center}
     \leavevmode\epsfxsize=10cm \epsfbox{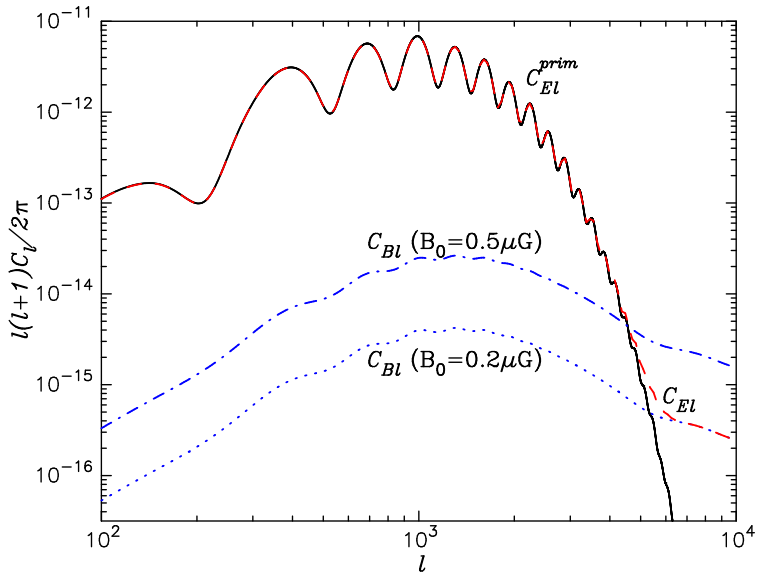}
\caption{The angular power spectra of $E$- and $B$-mode polarizations.
The dashed and dotted lines show the $E$- and $B$-mode spectra including
the Faraday mixing effect due to clusters at the observed frequency of
$10~{\rm GHz}$ for the magnetic field model of $B_0=0.2~\mu{\rm G}$,
respectively, while the dot-dashed line shows the $B$-mode spectrum for
the case of $B_0=0.5~\mu{\rm G}$ that corresponds to the value estimated
by Clarke et al. (2001) for a model of the uniform magnetic field.  For
comparison, the solid line shows the primary $E$-mode power spectrum in
the $\Lambda$CDM model.  \label{fig:clpol}}
 \end{center}
\end{figure}

In Figure \ref{fig:clpol} we show the angular power spectra of $E$- and
$B$-mode polarizations on the observed sky including the Faraday
rotation mixing effect due to clusters for the observed frequency of
$10{\rm GHz}$.  The dot-dashed line shows the $B$-mode spectrum for
$B_0=0.5\mu{\rm G}$ that corresponds to the value estimated from the
observational data for a model of uniform intracluster magnetic field
(Clarke et al. 2001), while the dotted line shows the result of
$B_0=0.2\mu{\rm G}$.  For comparison, the $E$-mode power spectra with
and without the secondary effect of $B_0=0.2\mu{\rm G}$ are also shown
by solid and dashed lines, respectively.  One can readily see that the
generated $B$-mode power spectrum has a peak with magnitude of $\sim
0.1~\mu{K}(B_0/0.1~\mu{G})(\nu_0/10~{\rm GHz})^{-2}$ at $l\approx 1000$.
On smaller scales of $l\simgt 6000$, the spectra of $E$ and $B$ modes
coincide, and this is similar as the secondary effects caused by the
peculiar velocity field of the ionized medium in the reionized universe
(\cite{Hu}; \cite{Liu2001}).  These results imply that, if the uniform
magnetic field has a strength of $\sim0.5~\mu{\rm G}$ and we can observe
at lower frequencies such as $\nu_0\simlt 5~{\rm GHz}$, the magnitude of
$B$-mode could be comparable with the primary polarization amplitude of
$\sim\mu{\rm K}$.

The detectability of the induced $B$ mode polarization by the sensitive
satellite mission Planck can be estimated following the method developed by
Zaldarriaga \& Seljak (1998).  The relative error on the overall
amplitude of the induced $B$ component $\beta$ can be estimated as
\begin{equation}
\frac{\Delta\beta}{\beta}=\sqrt{\frac{2}{f_{\rm sky}\sum_l
(2l+1)/(1+w^{-1}e^{l^2\theta_s^2}/C_{\tilde{B}l})^2}},
\end{equation}
where $f_{\rm sky}$ is the fraction of observed sky and we here fix
$f_{\rm sky}=0.8$. As for the detector noise and the beam width, we
assume $w^{-1}=4.72\times 10^{-16}$ and $\theta_{\rm s}=\theta_{\rm
fwhm}/ \sqrt{8\ln2}=3.71\times 10^{-3}$ for the $30$GHz channel on
Planck, where $\theta_{\rm fwhm}$ is the full-width at half-maximum
angle. Note that the $30$GHz channel we here consider is the lowest
frequency channel on Planck. From the dependence $\lambda_0^2$ of the
rotation measure, the magnitude of the $B$-mode angular power spectrum
at $30$GHz becomes smaller by $\sim 10^{-2}$ than the results estimated
in Figure \ref{fig:clpol}.  Accordingly, for the magnetic field of
$B_0=0.2~\mu{\rm G}$, we have $\beta/\Delta\beta\approx 0.14$.  This
means that this effect cannot be detected by the Planck mission for
$B_0=0.2\mu{\rm G}$. Conversely, to get $\beta/\Delta\beta\approx 1$ at
the $30$GHz channel, the uniform magnetic field component must have
$\sim 1.5~\mu{\rm G}$ strength, and this seems an unrealistic value.
However, we stress that even a null detection of this effect can set a
new constraint on the magnetic field strength.  It is also worth noting
that this signal would be detected more effectively by a ground based
experiments at lower frequency bands such as $\simlt 10~{\rm GHz}$
observing a small patch of the sky for a sufficiently long integration
time to reduce the noise per pixel.

%
%
\section{Discussions and Conclusions}
\label{Disc} In this paper, we have investigated the secondary effect on
the CMB polarization fields induced by the Faraday rotation effect of
the magnetic field in an intracluster hot plasma at a low redshift. To
illustrate the impact of this effect in the simplest way, we employed
the simple model that the magnetic field has a uniform field strength
such as $0.2~\mu{\rm G}$ across a cluster universally when the cluster
formed, which is consistent with observations of the rotation measures
in clusters (Clarke et al. 2001). As shown in Figure \ref{fig:averm},
our model can roughly reproduce the observed scatter of the rotation
measures as a function of the distance from the cluster center (Kim et
al. 1991; Clarke et al. 2001).  In this sense, since the Faraday
rotation effect on the angular power spectra of CMB polarizations comes
from the second moments of the rotation measure angle (see equation
(\ref{eqn:rmclB})), it is likely that our model can at least estimate a
correct magnitude of the effect, even though detailed shapes of those
spectra could be different for a realistic intracluster magnetic field
as discussed below.  We showed that the parity-odd $B$-mode angular
power spectrum is generated on the observed sky by the Faraday rotation
mixing effect as a new qualitative feature, when if the primary CMB
anisotropies includes the parity-even $E$-mode only such as predicted by
the standard inflation-motivated scenarios with pure primordial scalar
perturbations.  We estimated that the generated $B$-mode power spectrum
has a peak with amplitude of $\sqrt{l^2C_{lB}/2\pi}\sim 0.1~\mu{\rm K}
(B_0/0.1~\mu{\rm G}) (\nu_0/10~{\rm GHz})^{-2}$ at $l\approx 1000$ under
the plausible scenario of the cluster formation in the currently favored
$\Lambda$CDM model. It was also shown that the lowest frequency $30{\rm
GHz}$ channel of Planck can be used to set an  upper limit of
$B_0\simlt 1.5\mu{\rm G}$ for a uniform component of the intracluster
magnetic field.
Detection or even null detection of the predicted $B$-type polarization
will therefore be a new empirical tool to provide a calibration of an
uncertain magnetic field in a cluster, combined with measurements of the
gas distribution from the X-ray or SZ datasince.

It is known that there are other secondary sources on the CMB
polarization fields caused in the low redshift universe, and these
nonlinear effects generally induce the $B$-type polarization pattern
from the coupling with the primary $E$-mode.  The main source is the
gravitational lensing effect of the large-scale structure, leading to
the $B$-type polarization that has a peak of $\sim 0.3~\mu{\rm K}$ at
$l\approx 1000$ and then a power of $\simlt 0.01~\mu{\rm K}$ at $l\simgt
5000$ for the simliar $\Lambda$CDM model as considered in this paper
(\cite{ZS98}).  One of authors (N.S.) has quantitatvely shown that the
secondary effect due to the peculiar motion of ionized medium in the
large-scale structure also generates the $B$-mode polarization with
$\sim 0.01~\mu{K}$ at $100\simlt l \simlt 10^4$ for a realistic patchy
reionization model of the universe (\cite{Liu2001}). From these results,
if the uniform magnetic field has a larger strength of $B\sim
0.5~\mu{\rm G}$ as a common feature of clusters and we observe at lower
frequencies such as $\nu_0\simlt 5~{\rm GHz}$, the generated $B$-mode
amplitude due to the Faraday rotation effect would be comparable with
the primary polarization amplitude and larger than the other secondary
signals at $l\simgt1000$, although observations at such low frequencies
would suffer from large foreground contamination of the syncrhrotron
radiation from our galaxy (\cite{Tegmark}).  Even if the uniform
magnetic field is not as strong as $0.5~\mu{\rm G}$, the sensitive multi
low-frequency measurements will allow one to separate this signal from
primary and those other secondary signals in principle thanks to the
frequency dependence of $\propto \nu_0^{-4}$ on $C_{Bl}$, which is
analogous to measurements of secondary temperature fluctuations by the
SZ effect.

There are possible contributions to affect our results that we have
ignored in this paper.  The intracluster tangled magnetic fields, as
implied by the rotation measures of some sources inside the Coma cluster
(\cite{Kim90}), would provide contributions to our results in addition
to the contribution due to the uniform field. However, the coherent
scale of tangled fields in a cluster is still unknown. The recent
complete measurements of the rotation measures (Clarke et al. 2001)
imply the $1-10~\mu{\rm G}$ strength of the tangled field under the
assumption of the coherent scale of $10-100~{\rm kpc}$.  It is also
known that the strong radio halos embedded within a (cooling flow)
cluster are pervaded by stronger magnetic fields on smaller scales such
as $\sim 50~{\mu G}$ for Hydra A (\cite{Taylor}; \cite{Taylor01}).
Although the simplest tangled-cell model with a constant coherence
length $l$ has often been considered in the literature, the possible
important mechanism to affect our results is the deporlarization effect
(\cite{Tribble}), leading to the cancellations to some extent between
the Faraday rotations in independent cells with random orientations of
the magnetic fields within a cluster.  The finite beam size will also
lead to the artificial cancellations between the Faraday rotations of
cells covered within one beam. From these considerations, the
depolarization effect may lead to an intuitive result that a feature of
the $B$-mode spectrum can be sensitive to the characteristic angular
scale, $l_{c}$, originating from the coherent length.
This investigation is now in progress, and will be presented elsewhere
(Ohno et al. 2001).  Recently, it has been suggested that
outflow or jet acitivities of quasars or black holes leave behind an
expanding magnetized bubble in the itergalactic medium (IGM) resulting
in the possibility that the IGM is filled by magnetic fields to some
extent (\cite{Furlan}; \cite{Kronberg01}). Even thgouh the IGM field has
a smaller strength than $\sim 10^{-9}{\rm ~G}$ on ${\rm Mpc}$ scales, a
large volume coverage of the ionized plasma might lead to non-negligible
contribution connected to the reionization history of the universe.

Finally, we comment on implications of our results to the nonthermal
processes in a cluster.  The observations of diffuse synchrotron halos
and nonthermal hard X-ray emission have implied the existence of
nonthermal relativistic electrons in a cluster (e.g.,
\cite{Rephaeli99}).  These electrons are produced by the shock
acceleration in an intracluster medium, and it will allow us to derive
additional information on the physical conditions of the intracluster
medium environment, which cannot be obtained from the thermal plasma
emission only.  The intracluster magnetic field should then play an
important role to these nonthermal processes. In particular, it is being
recognized that the nonthermal process can be a unique probe of
measuring the dynamically forming clusters before thermalization of the
intracluster gas.  Recently, Waxman \& Loeb (2000) pointed out that the
synchrotron radiation from those forming clusters could be foreground
sources of the CMB temperature fluctuations at low frequency
observations such as $10{\rm ~GHz}$.  Those clusters could also produce
the high energy gamma ray emission through the inverse Compton
scattering of CMB photons by relativistic electrons and thus be
identified as {\it gamma ray clusters} by the future sensitive telescope
{\em GLAST}, which are difficult to detect by X-ray or optical surveys
because of their extended angular size of about $\sim 1^\circ$
(\cite{Totani}).  Based on these ideas, it will be very interesting to
investigate cross-correlations between the synchrotron radiation or the
gamma ray sources and the Faraday rotation signal in CMB polarization
map. This investigation is expected to provide complementary constraints
on the nonthermal physical processes in the forming clusters.

\section*{Acknowledgments}
M.~T. is grateful to Prof. T.~Futamase and O.~Aso.  This work was
initiated from the conversation with Prof. T.~Futamase about the
original idea of O.~Aso. They and M.~Hattori have also done the similar
work as this paper.  M.~T. also thanks M. Chiba, M.~Hattori, N.~Seto and
H.~Akitaya for useful discussions and comments.  We thank U. Seljak and
M. Zaldarriaga for their available CMBFAST code.  M.~T. acknowledges
support from the Japan Society for Promotion of Science (JSPS) Research
Fellowships for Young Scientist. This work was supported by Japanese 
Grant-in-Aid for Science Research Fund of the Ministry of Education, 
Science, Sports and Culture Grant Nos. 01607 and 11640235, and Sumitomo 
Fundation.

\end{document}